# Frequency-multiplexed storage and distribution of narrowband telecom photon pairs over a 10-km fiber link with long-term system stability

Ko Ito[1], Takeshi Kondo[1], Kyoko Mannami[1], Kazuya Niizeki[1,3], Daisuke Yoshida[1,3], Kohei Minaguchi[1], Mingyang Zheng[2], Xiuping Xie[2], Feng-Lei Hong[1], and Tomoyuki Horikiri[1,*]

[1]*Department of Physics, Graduate School of Engineering Science, Yokohama National University, Yokohama 240-8501, Japan*
[2]*Jinan Institute of Quantum Technology, Jinan 250101 Shandong, China*
[3]*LQUOM Inc., Yokohama 240-8501, Japan*
*\*horikiri-tomoyuki-bh@ynu.ac.jp*

**Abstract:** The ability to transmit quantum states over long distances is a fundamental requirement of the quantum internet and is reliant upon quantum repeaters. Quantum repeaters involve entangled photon sources that emit and deliver photonic entangled states at high rates and quantum memories that can temporarily store quantum states. Improvement of the entanglement distribution rate is essential for quantum repeaters, and multiplexing is expected to be a breakthrough. However, limited studies exist on multiplexed photon sources and their coupling with a multiplexed quantum memory. Here, we demonstrate the storing of a frequency-multiplexed two-photon source at telecommunication wavelengths in a quantum memory accepting visible wavelengths via wavelength conversion after 10-km distribution. To achieve this, quantum systems are connected via wavelength conversion with a frequency stabilization system and a noise reduction system. The developed system was stably operated for more than 42 h. Therefore, it can be applied to quantum repeater systems comprising various physical systems requiring long-term system stability.

# I. INTRODUCTION

The quantum internet has the potential to be used in various applications such as distributed quantum computation, cloud quantum computing [1], world clocks [2], and ultralong baseline interferometry [3]. In the quantum internet system, quantum repeaters are indispensable for sharing the entanglement of the quantum states (qubits) between adjacent repeater nodes to realize long-distance quantum communication. Quantum repeaters can be implemented in either an optical-fiber-based or a satellite-based [4] quantum communication system. Furthermore, a quantum internet using hybrid quantum communication between satellites and optical fibers is coming to fruition [5–8].

Quantum memory, which stores a quantum state for later retrieval, is a key component of the quantum repeater. Various materials working at different wavelengths have been studied for quantum memory applications [9–18]. Using wavelength conversion as an intermediate step [17–20], the diversity of a network consisting of various materials can be acquired. Atomic frequency comb (AFC) [21] quantum memories based on rare-earth-doped materials are promising for multiplexed quantum communication, and are essential for improving the entanglement generation rate of quantum communication [13,14]. Rare-earth-doped materials have gained considerable attention owing to the high coherence of their donors and their multiplicity by controlling the inhomogeneous width. The AFC method accepts single photons from collective donor ions, thus exhibiting high absorption and storage efficiency and enabling high retrieval efficiency through rephasing. Among these, $Pr^{3+}:Y_2SiO_5$ (Pr:YSO) has demonstrated a relatively long coherence time, high multimode capacity, and high efficiency [13,22]. The AFC and memory transition wavelength of this material is approximately 606 nm; therefore, the transmitted telecommunication wavelength photons need to be wavelength-converted before being stored in the memory. This wavelength conversion can be realized based on sum-frequency generation (SFG): the frequency of the memory photon (606 nm) is the sum of the telecom photon (1.5 μm) and a wavelength conversion pump laser (1.0 μm). Owing to the narrow bandwidth of the AFC, the frequency of the photons produced by SFG must be stabilized at the AFC frequency. Therefore, all the components, including telecom entangled-photon sources and wavelength conversion pump lasers, must also be stabilized in addition to the quantum memory control lasers for AFC generation, which is the reference for these lasers.

A typical quantum repeater node has at least two quantum memories and, if necessary, after both quantum memories are loaded, photons are regenerated for quantum-entanglement swapping [23]. Indistinguishability is required for successful entanglement swapping. The spectra (center frequency and linewidth) as well as the timing of incidence on the Bell measurements for swapping must be matched. For example, because the wavelength conversion pump laser and the quantum memory control laser can be the same for the two memories used in the intranode swapping operation, the problem of spectral quantum interference between the regenerated photons from the two quantum memories can be addressed. Furthermore, by sending a laser (telecommunication wavelength laser in the case of optical fiber transmission) that is stabilized with respect to the frequency of an optical frequency comb in one node, which is the frequency reference, frequency coordination between several repeaters is possible. Using this scheme, a multinode quantum repeater through entanglement swapping between remote quantum devices can be developed if the frequency matching between the elements [24] is stable over time.

To establish a system that can perform entanglement swapping operations at quantum repeater nodes, it is necessary to combine remote quantum devices, that is, a frequency-multiplexed telecom photon source and a frequency-multiplexed quantum memory. Photons emitted from a frequency-multiplexed telecom photon source will need to be sent through a quantum communication channel consisting of a long optical fiber, wavelength-converted, and stored in a frequency-multiplexed quantum memory. Until now, there has been no example of coupling telecommunication wavelength two-photons into a frequency-multiplexed memory through wavelength conversion after long optical fiber transmission. This can be attributed to the difficulty of implementing wavelength conversion with a high signal-to-noise ratio (SNR) under low photon rates after long-distance transmission as well as the difficulty of frequency coordination between systems.

In this study, we place an optical frequency comb as an optical frequency reference in the repeater node to enable the frequency tuning of the involved telecommunication wavelength photons, wavelength conversion (pump laser), and quantum memory control laser. We also introduce a noise reduction system to obtain a high SNR for the wavelength conversion system. Using these systems, we observe photon signals by storing and retrieving photons from a narrow linewidth two-photon source [13,25,26] in a fixed-time quantum memory following fiber transmission over 10 km. Because various physical systems can be used in the physical layer that constitutes the quantum internet, it is highly likely that the system will be a wavelength-conversion-mediated system for quantum communication in both satellites and optical fibers. This study is a significant step in that direction.

## II. RESULTS

### A. Experimental setup

In this section, we describe the overall setup of the experiment, as illustrated in Fig. 1. The experiment can be divided into the following parts: a telecom two-photon source [two-photon comb (TPC) [25]], wavelength conversion, quantum memory, noise reduction system, and an additional frequency stabilization system for the highly efficient coupling of the TPC and the quantum memory. In the following paragraphs, we will describe each component.

The TPC consists of a periodically poled lithium niobate (PPLN) chip for generating two photons through degenerate spontaneous parametric down-conversion (SPDC) and its surrounding cavity. The wavelength of an external-cavity diode laser (ECDL) at 1514 nm was converted to 757 nm through second-harmonic generation (SHG) and used as the pump laser for SPDC. The generated photon pair had a cluster width [26] of approximately 500 GHz, a linewidth of 7.1 MHz in the 1514 nm region, and a free spectral range (FSR) of 117.2 MHz.

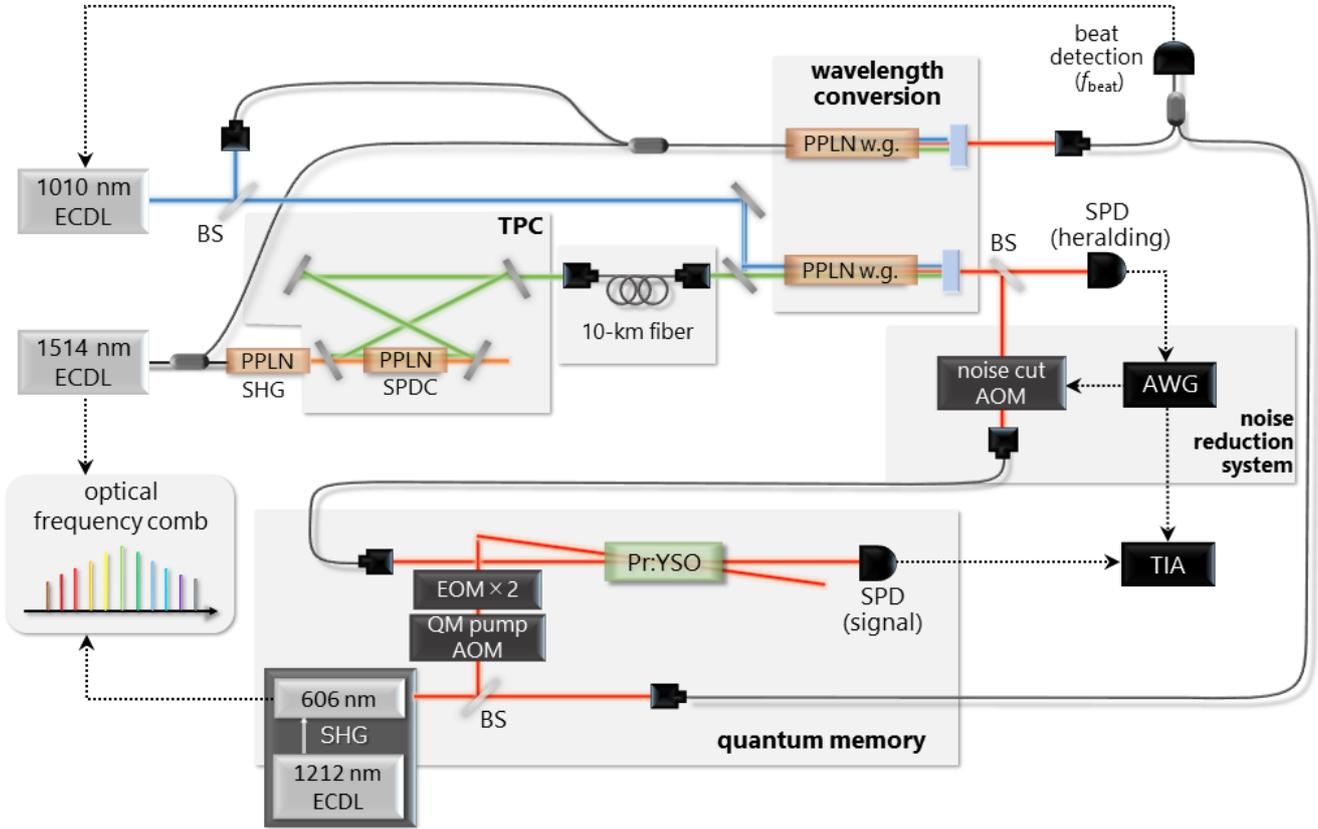

FIG. 1. Schematic of the experimental setup. The experiment comprises a two-photon comb, quantum memory, wavelength conversion system, noise reduction system, and frequency stabilization system. The SHG light (757 nm) from the TPC pump laser (1514 nm) caused SPDC inside the cavity at a pump power of 25 mW, generating a degenerate photon pair (1514 nm). After transmission over a 10-km length optical fiber, the pair was converted to a quantum memory wavelength (606 nm) by SFG using a PPLN waveguide. The wavelength-converted photons were split into two using a BS, and one side was detected by a SPD and used as a heralding photon. The other side was stored in quantum memory after transmission through the AOM used in the noise reduction system, and the regenerated photons were detected by a SPD. The time difference between the detection on the heralding side and detection after the quantum memory was measured by TIA. The frequency of all lasers and the TPC cavity was locked to the appropriate frequency by the frequency stabilization system.

A Pr:YSO crystal with a doping ratio of 0.05% was used for the quantum memory. The crystal was placed inside a cryostat (4 K) and had an inhomogeneous broadening of approximately 10 GHz, as depicted in Fig. 2(a). Moreover, a transparent region (pit) was created in the crystal, where an AFC (comb-shaped absorption lines) was created for photon storage and retrieval. The frequency and timing of the quantum memory control laser (606 nm), which was generated through the SHG of an ECDL at 1212 nm, were adjusted by the RF signal applied to the acousto-optic modulator (AOM) installed in the path. Initially, a small absorption region was formed by sweeping around the frequency of $1/2g$-$1/2e$ transition of $^3H_4(0)$-$^1D_2(0)$, as depicted in Fig. 2(a), in the range of ±9 MHz. Next, a sharp peak was formed in the transparent region by injecting a single-frequency light with a frequency of $5/2g$-$5/2e$ transition into the crystal (this process is referred to as "burn back" [22]). The AFC was formed by repeating this process at the peak spacing of each comb (1.15 MHz in this experiment). The reference frequency in the region where this AFC is generated [$\nu_{AFC}$, the zero point in the inset of Fig. 2(b)] is the quantum memory control laser frequency $\nu_{QM}$ with a constant offset applied by an AOM ($f_{QMpumpAOM}$ = 164.3 MHz). The frequency matching between the TPC and AFC is performed as follows: the TPC laser (1514 nm) is locked to an optical frequency comb, which is a frequency reference in this study that synchronizes with the global positioning system signals (GPS comb). The frequency of the quantum memory control laser is also locked to the GPS comb. Therefore, when the TPC laser is converted to 606 nm through wavelength conversion, a beat measurement is made, with the quantum memory control laser responsible for AFC generation, and the feedback is applied to the wavelength conversion laser

frequency to keep the beat frequency constant. (Further details are provided in Section II.B.)

In this study, we also performed the frequency multiplexing of the quantum memory using two electro-optic modulators (EOMs) after transmitting the light through an AOM [27]. We applied a signal of frequency $\nu_{FSR}$ (117.2 MHz) corresponding to the FSR of the TPC to the first EOM, and a signal of $5 \times \nu_{FSR}$ (586.0 MHz) to the second EOM. By adjusting the amplitude of the RF signal applied to the EOMs, 0, ±1, and ±2 order peaks were formed for each EOM, resulting in the generation of AFCs for 25 frequency modes. Figure 2(b) depicts the generated AFCs. In this study, the optical power of the quantum memory control laser was 2.3 mW.

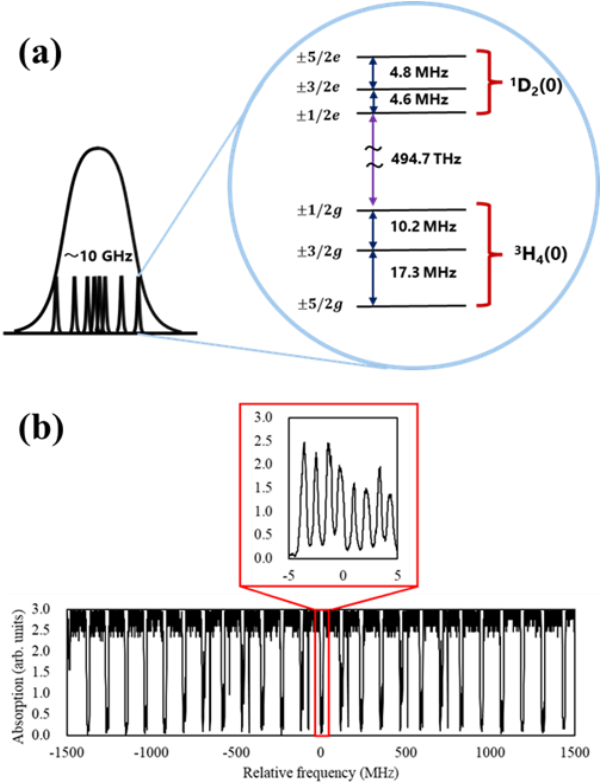

FIG. 2. Quantum memory. (a) Inhomogeneous broadening of Pr:YSO (left) and energy diagram (right). (b) This panel shows the entire 25-frequency-mode quantum memory; the enlarged panel shows a magnified view of one of the frequency modes.

Owing to the gap between the telecommunication wavelength photon (1514 nm) generated by the TPC and the wavelength of the Pr:YSO quantum memory (606 nm), the photon from the TPC could not be directly stored in the quantum memory. Therefore, we converted the telecommunication wavelength to the quantum memory wavelength (606 nm) by performing SFG using a wavelength conversion pump laser (1010 nm) in a PPLN waveguide after the long optical fiber transmission of photons generated in the TPC. Following wavelength conversion, only photons with wavelengths of approximately 606 nm were extracted using two bandpass filters and two dichroic mirrors. This experiment was performed using a wavelength conversion pump laser with a power of 140 mW and a quantum wavelength conversion efficiency of 55.8%. Limited by the phase-matching bandwidth of the PPLN waveguide, the full width at half maximum (FWHM) bandwidth of this wavelength conversion system was approximately 40 GHz. The wavelength-converted photon pair was split equally by a beam splitter (BS), and one side was detected as a heralding photon by a single-photon detector (SPD). The other side was detected by an SPD after a single photon from the TPC had been stored and reemitted by the frequency-multiplexed quantum memory. The time difference between the detection of the heralding photon and the other photon was measured using a time interval analyzer (TIA).

### B. Frequency stabilization system

To couple wavelength-converted photons with quantum memory, it is necessary to precisely match the wavelength of the wavelength-converted photons with that of the quantum memory control laser. We achieved the desired frequency matching by developing a wavelength conversion system for monitoring, which is different from that used for the photons. This is described as follows (see also Fig. 3.). The frequencies of the TPC pump laser at approximately 1514 nm ($\nu_{photon}$) and the master of the quantum memory control laser at approximately 1212 nm ($\nu_{QMmaster}$) were offset-locked using a delay line [24,28–30] into an optical frequency comb that synchronizes with the GPS comb. Further, the resonant frequency of the TPC cavity was locked using the Pound–Drever–Hall method with a TPC pump laser [25]. The temperature of the PPLN is well stabilized within ~1 mK. In general, the resonant frequency of the TPC cavity is well locked to the TPC photon laser. A wavelength conversion system was developed for monitoring, and feedback was returned to the wavelength conversion pump laser by offset-locking such that the frequency of the beat signal between the 606 nm laser light for monitoring ($\nu_{monitor}$) and the frequency of the quantum memory control laser ($\nu_{QM} = 2 \times \nu_{QMmaster}$) became constant. (The detailed scheme is described in the caption of Fig. 3).

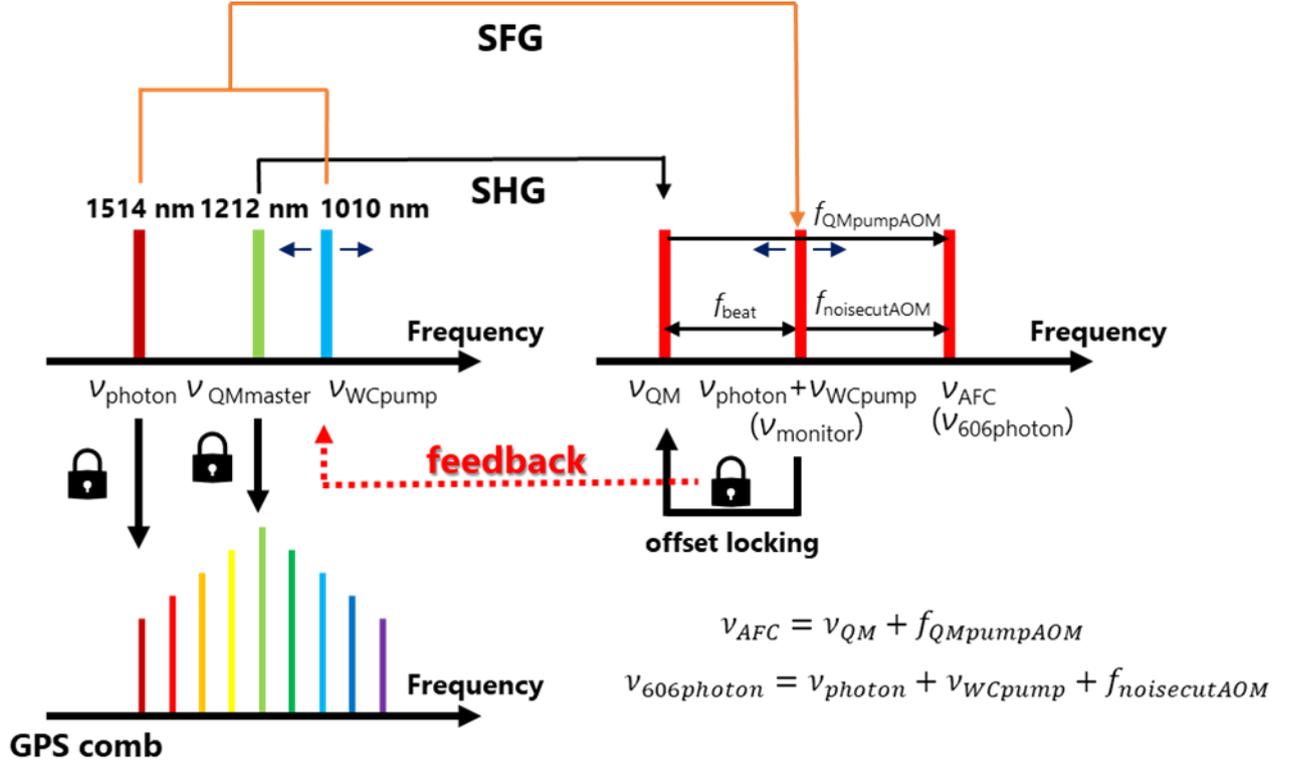

FIG. 3. Frequency locking diagram.
First, the TPC pump laser (1514 nm) and the master of the quantum memory control laser (1212 nm) are offset-locked to the GPS comb. Thereafter, the wavelength-converted light for monitoring is offset-locked to the quantum memory control laser (606 nm). Offset-locking is achieved through feedback to the wavelength conversion pump laser (1010 nm) for obtaining the frequency match of telecom TPC photons with the AFC quantum memory. Because the photon before entering the quantum memory is a wavelength-converted photon generated from the TPC and passed through the noise-cutting AOM (which will be discussed later) of modulation frequency ($f_{\mathrm{noisecutAOM}}$), its frequency ($\nu_{\mathrm{606photon}}$) is expressed as follows:

$$\nu_{\mathrm{606photon}} = \nu_{\mathrm{monitor}} + f_{\mathrm{noisecutAOM}},$$

where $\nu_{\mathrm{WCpump}}$ is the frequency of the wavelength conversion pump laser. Conversely, because quantum memory is generated by modulating the quantum memory control laser with an AOM (modulation frequency, $f_{\mathrm{QMpumpAOM}}$), the frequency at which the atomic comb structure ($\nu_{\mathrm{AFC}}$) is actually created is expressed as follows:

$$\nu_{\mathrm{AFC}} = \nu_{\mathrm{QM}} + f_{\mathrm{QMpumpAOM}} = 2 \times \nu_{\mathrm{QMmaster}} + f_{\mathrm{QMpumpAOM}}.$$

Because the 606-nm light for monitoring is offset-locked at frequency $f_{\mathrm{beat}}$ to the quantum memory control laser, we obtain the following expression:

$$\nu_{\mathrm{monitor}} = \nu_{\mathrm{QM}} + f_{\mathrm{beat}}.$$

Therefore, we obtain

$$f_{\mathrm{QMpumpAOM}} = f_{\mathrm{beat}} + f_{\mathrm{noisecutAOM}}$$

by adjusting the frequency of $\nu_{\mathrm{WCpump}}$ using the feedback information ($f_{\mathrm{beat}}$) so that

$$\nu_{\mathrm{606photon}} = \nu_{\mathrm{AFC}}$$

and relative frequency matching is achieved.
In the experiment, the values of $f_{\mathrm{QMpumpAOM}}$, $f_{\mathrm{beat}}$, and $f_{\mathrm{noisecutAOM}}$ were set as 164.3, 83.4, and 80.9 MHz, respectively.

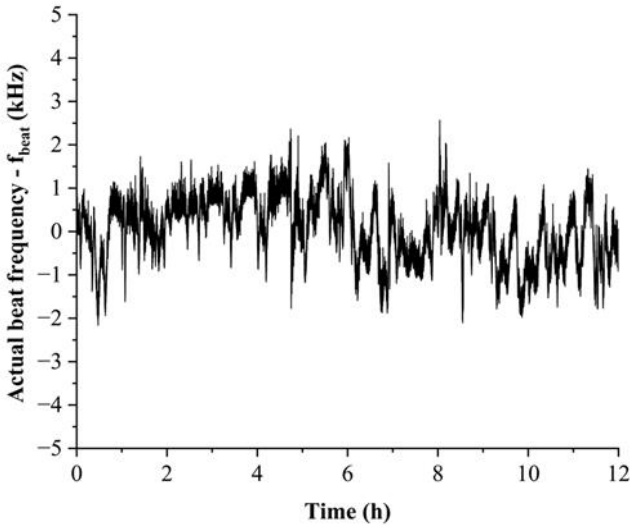

FIG. 4. Result of the relative frequency stabilization. In a 12 h measurement, the relative frequency drift is maintained below 5 kHz.

Figure 4 presents the experimental results of the relative frequency-locking. The drift of the beat frequency between $\nu_{\mathrm{QM}}$ and $\nu_{\mathrm{monitor}}$ for monitoring is illustrated. The frequency drift was suppressed within 5 kHz over 12 h, which is sufficient for accurate frequency matching between the wavelength-converted photons and AFCs.

### C. Noise reduction system

In this study, noise mainly originates from the strong wavelength conversion pump laser [31–33]. The TPC photons retrieved from the quantum memory are buried in the noise without a noise reduction system; therefore, we separated the retrieved photons from the noise through an installed noise reduction system [34]. The noise originating from the wavelength conversion pump laser was removed by a fast shutter consisting of an AOM. Figure 5 depicts the timescale of the entire experiment. The cycle is repeated throughout the experiment. During the photon transmission time, the AOM shutter placed after the wavelength conversion system is operated as follows. In particular, we closed the AOM shutter based on the timing of the arrival of the heralding photons to the SPD. By shuttering the light after wavelength conversion using the time difference until the photons are generated as echoes, we could separate the signal photons in the quantum memory from the noise originated from the continuously operating strong wavelength conversion pump laser. The combination of AFC and time filtering results in filtering in the frequency domain as well because time filtering only detects retrieved photons that have been absorbed and retrieved by the AFC.

Noise originates mainly from the wavelength conversion pumping laser. The pump laser (of wavelength approximately 1 µm) produces large noise at the telecommunication wavelength owing to SPDC and Raman scattering inside the nonlinear medium when visible signal photons are converted into telecommunication wavelengths through difference frequency generation [35]. In contrast, in our case, the SFG is used for converting telecom signal photons into visible light; the pump laser at approximately 1 µm generates the same noise at the telecommunication wavelength. This noise is also upconverted to quantum memory wavelength in the wavelength conversion device, causing the noise in the visible quantum memory wavelength to be less than the originally generated telecommunication wavelength noise. In our case, with a phase-matched bandwidth of approximately 40 GHz, and including the wavelength conversion efficiency (<1), the noise was approximately 40 kcps at a pump power of 140 mW. Furthermore, Fig. 6 indicates that the noise can be reduced by two orders of magnitude with a noise filtering system.

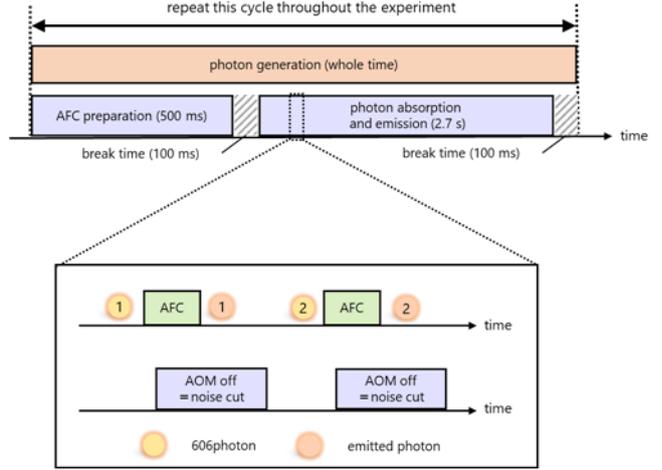

FIG. 5. Timescale of the experiment. Photons are constantly generated from the TPC. The quantum memory experiment is divided into the time to prepare the AFC and the time to transmit photons. During the photon transmission time, the AOM shutter reduces noise from the cavity and wavelength conversion pump lights.

### D. Experimental results

Figure 6 presents the experimental results of the time difference between the heralding photon and the photon regenerated from the quantum memory measured using the TIA. The AFC was created for 25 frequency modes. TPC photons were transmitted over a 10 km optical fiber. The time resolution of the TIA was 0.128 ns and the measurement time was 12 h. The peaks on the left side represent the transmitted photons outside the inhomogeneous broadening of Pr:YSO owing to the fact that while the bandwidth of the wavelength conversion is approximately 40 GHz, the inhomogeneous broadening width of Pr:YSO used in this study is approximately 10 GHz. Therefore, some of the wavelength-converted photons are transmitted outside the inhomogeneous broadening. The peaks on the right side represent the photons regenerated from the AFC. Defining $S$ as the total number of counts including noise and $N$ as the number of noise counts, the signal-to-noise ratio (SNR) can be expressed as SNR = $(S - N)/N$. The signal count $S$ at the peak is 74 counts and the SNR is 1.4. In this result, small peaks were observed at approximately 150 ns, which is caused by the slow light effect. The slow light effect [36,37] is caused by the gradient of the absorption peaks, that is, the gradient of the refractive index. The retrieval time of the echoed photons observed in this experiment (the appearance time of the peaks on the right) is the sum of the delay time owing to the slow light effect and storage time in the quantum memory.

We also used this system to extract a few frequency modes by replacing the long optical fiber with a 5 m fiber. Figure 7 depicts the situation when the 5- or 1-frequency mode is generated. The vertical axis represents the SNR, and the time on the horizontal axis is the same as Fig. 6. The measurement times were 12 and 42 h, respectively, and the time resolution of the TIA was set at 0.128 ns. These results indicate that the SNR at each peak is 0.87 and 0.17, respectively. Because a frequency-multiplexed two-photon source is utilized in this study, a large frequency multiplicity leads to a high SNR as the probability of two pairs generating simultaneously in the same frequency mode decreases. Therefore, the pump power of the TPC can be increased when compared with the 1-frequency mode case.

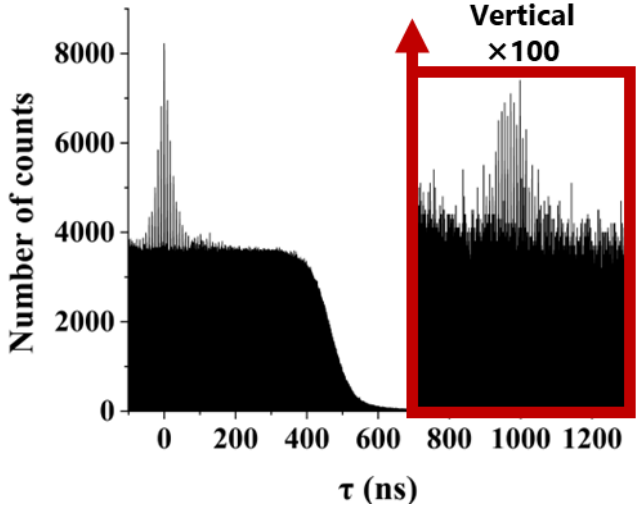

FIG. 6. Results of a 10-km optical fiber transmission experiment. This experiment was performed with a TIA time resolution of 0.128 ns. Peaks derived from the retrieved photon from the quantum memory can be observed in the noise-cut region (right side of the figure).

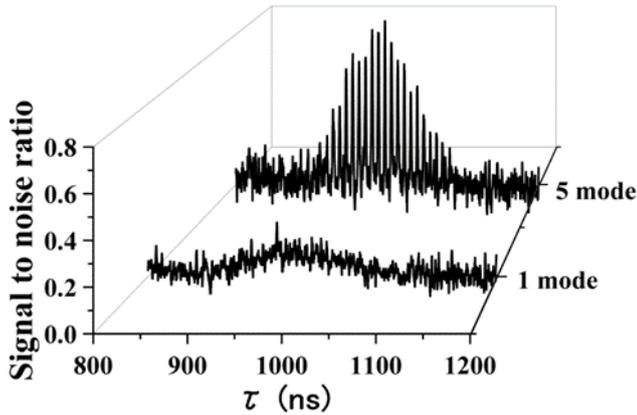

FIG. 7. Experimental results pertaining to the 1-mode and 5-mode. The time resolution of the TIA is 0.128 ns, and the adjacent average is considered for 10 bins (= 1.28 ns). These data are normalized by the noise floor.

## III. DISCUSSION

In this study, we generated 1–25 wavelength channels in our quantum memory and demonstrated 10 km optical fiber transmission of frequency multiplexed TPC. In actual quantum communication, by creating a situation in which each wavelength channel can be separately detected, indistinguishability [23,38] can be obtained in the Bell measurements and an improvement in the quantum-entanglement generation rate between quantum memories can be achieved with the merit of frequency multiplexing. It will be necessary to acquire the ability to identify each wavelength channel and perform quantum-entanglement swapping. Previous studies have investigated quantum interference between multiple wavelengths, including wavelength- to spatial-mode switching [38] and wavelength-to-time conversion to discriminate by varying the AFC retrieval time for each channel [39]. This indicates that wavelength-division-multiplexed (WDM) entanglement swapping for quantum repeaters is possible. For a two-photon source located in a memory-equipped repeater node, the photon generates a heralding signal for the entanglement swapping operation through optical Bell measurement at the midpoint of the optical fiber transmission [23,40], where the two above-mentioned methods can be applied. In this study, it is assumed that quantum-entangled two-photon generation is performed at a midpoint separated from the quantum memory [41,42], and that the photons are transmitted to each quantum memory on both the sides, which requires wavelength distinguishability in the quantum memory. When employing AFC quantum memories, as in this study, the AFC comb's peak spacing of 25 channels can be changed to identify the channels using their retrieval time[43]. It is possible to increase the number of frequency multiplexing channels to approximately 50, utilizing the inhomogeneous broadening width of Pr:YSO. This would increase the excitation power of the entangled photon source, as described in the Results, which would increase the SNR and improve the resulting quantum-entanglement fidelity.

In this study, we have successfully measured two-photon correlation after fiber transmission over 10 km. However, the small signal-to-noise ratio is an obstacle in the demonstration of quantumness including high-fidelity entanglement. Future improvements will include the use of two-photon sources with higher spectral brightness, which can lead to higher signal-to-noise ratio, such as a nondegenerate two-photon source, which can narrow the two-photon generated cluster width by two orders of magnitude or more [26,44] and, thereby, improve the rate of signal photons that couple with the AFC.

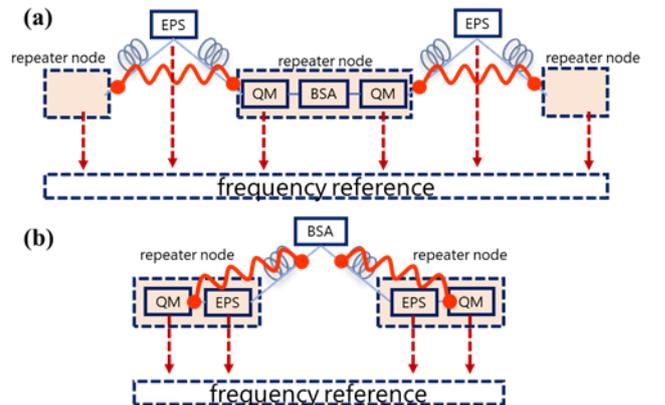

FIG. 8. Available schemes for entanglement generation. (a) midpoint source, (b) meet-in-the-middle.

The photon source, quantum memory, and wavelength conversion system demonstrated in this study can be stably coupled for long periods of time using a frequency stabilization system. The present coupling system of remote quantum devices through the frequency stabilization system will also enable a connection to other physical systems. If the system is to be connected to other quantum memory systems [15,16,19] as an end node of the quantum internet, the propagating photon must acquire indistinguishability [20] by wavelength conversion and frequency stabilization. This is because a quantum repeater through Bell measurements among hybrid physical systems becomes possible only when indistinguishability is achieved. Moreover, because indistinguishability is expected to be mediated by wavelength conversion [20], the method proposed in our study can be used.

In this study, an optical fiber quantum repeater is assumed because a telecommunication wavelength quantum photon source is used, but the developed coupled system can also be used in a hybrid quantum internet, where a satellite-based quantum repeater is combined. This is because wavelength conversion is necessary to connect the photon wavelengths used for satellite-to-ground quantum communication with the ground-side quantum internet nodes. In terms of feasibility, GPS combs can potentially be used as a frequency reference for the frequency stabilization of quantum devices onboard satellites for quantum repeaters [6–8].

There are variations in entanglement generation between the elementary links towards quantum repeaters. In the midpoint source scheme [41,42], two entangled photons of a telecommunication wavelength are sent to repeater nodes on the left and right, respectively, as in this study [Fig. 8(a)]. At these points, they are wavelength-converted to quantum memory wavelengths and absorbed and stored in quantum memories. This method enables entanglement sharing at a higher rate when compared with conventional methods [41]. However, the drawback of this method is that it cannot generate heralding signals when using absorptive quantum memories. Therefore, a method such as nondestructive photon arrival detection [42,45] is required to be implemented just before the memory, which is technically difficult at present. In contrast, in the meet-in-the-middle scheme [Fig. 8(b)], the entangled photon source is installed in the same node as the quantum memory. In this case, a single photon from the generated two-photon (degenerate telecommunication wavelength, as used in this study) is converted to the quantum memory wavelength on the spot and is absorbed and stored to the quantum memory. The other photon is measured at an intermediate station through an optical fiber to generate an entanglement-heralding signal in the elementary link [46]. Thus, this scheme overcomes the drawback that an absorptive quantum memory itself cannot generate heralding signals. As the wavelength conversion efficiency of photons stored in a quantum memory can be more than 50%, there is no significant loss caused by this. It is also possible to generate one of the two photons directly at the quantum memory wavelength, in which case wavelength conversion itself can be avoided.

In addition, the present wavelength conversion can also be incorporated into the meet-in-the-middle case. As studied in the frequency identification among frequency multiplexed system [43], Pr:YSO can be responsible for the frequency identification for the incoming telecom photons at the intermediate station. Therefore, the present efficient wavelength conversion system and the frequency identification system utilizing AFC generated in Pr:YSO at the intermediate station will enable frequency-multiplexed BSM.

To connect multiple repeater nodes, it is necessary to transmit a telecommunication wavelength laser between nodes in the case of an optical fiber. Because the loss of the telecommunication wavelength signal is 0.2 dB/km, assuming the typical distance of repeater nodes is 50 km, the loss is 10 dB, which means that the laser can be transmitted to multiple nodes comprising the quantum internet without any problem. In addition, because the stabilizing laser is a classical light, it is relatively easy to transmit over long distances by inserting a fiber amplifier in the middle of the transmission. For example, it is possible to send a 1514 nm laser that is absolute frequency-locked to a GPS optical frequency comb to a remote node through an optical fiber. The 1212 nm laser, which is a fundamental laser of the quantum memory control laser, can be frequency-stabilized under the same reference because it can be used as a reference for another optical comb installed within that node.

In this manner, our optical frequency-comb-based photon source–quantum memory coupling system can be deployed globally with a unified frequency standard.

In this study, we succeeded in coupling photons generated by a telecommunication-wavelength TPC to a frequency-multiplexed quantum memory through wavelength conversion after long optical fiber transmission. The developed system is promising in the context of the future quantum repeater architecture. The next step will be to demonstrate a frequency-multiplexed quantum repeater and expand it to multiple repeater nodes using the developed system.

## ACKNOWLEDGMENTS

We thank Qiang Zhang, Kazumichi Yoshii, Ippei Nakamura, Shuhei Tamura, Tomoki Tsuno, Takuto Miyashita, Yuma Goji, and Ryo Onozawa for their support in the experiment. Tomoyuki Horikiri also acknowledges members of the Quantum Internet Task Force, which is a research consortium set with the task of realizing the quantum internet, for comprehensive and interdisciplinary discussions of the quantum internet. We acknowledge funding from SECOM foundation, JST PRESTO (JPMJPR1769), JST START (ST292008BN), JSPS KAKENHI Grant Number JP20H02652, NEDO (JPNP14012), and JST Moonshot R&D Grant Number JPMJMS226C.

.